\documentclass[preprint]{aastex}

\shorttitle{Kepler EB  Catalog: Classification with LLE}
\shortauthors{Matijevi\v c et al.}

\begin{document}

\title{{\it Kepler} Eclipsing Binary Stars. III. Classification of Kepler Eclipsing Binary Light Curves with Locally Linear Embedding}

\author{Gal Matijevi\v c}
\affil{Faculty of Mathematics and Physics, University of Ljubljana, Jadranska 19, 1000 Ljubljana, Slovenia}
\email{gal.matijevic@fmf.uni-lj.si}

\author{Andrej Pr\v sa}
\affil{Dept. of Astronomy and Astrophysics, Villanova University, 800 E Lancaster Ave, Villanova, PA 19085, USA}
\email{andrej.prsa@villanova.edu}

\author{Jerome A.\ Orosz and William F.\ Welsh}
\affil{Department of Astronomy, San Diego State University, 5500 Campanile Dr., San Diego, CA 92182, USA}

\author{Steven Bloemen}
\affil{Instituut voor Sterrenkunde, KU Leuven, Celestijnenlaan 200 D, B-3001 Leuven, Belgium}

\and
\author{Thomas Barclay}
\affil{NASA Ames Research Center/BAER Institute, Moffett Field, CA 94035, USA}

\begin{abstract}
We present an automated classification of 2165 \textit{Kepler} eclipsing binary (EB) light curves that accompanied the second \textit{Kepler} data release. The light curves are classified using Locally Linear Embedding, a general nonlinear dimensionality reduction tool, into morphology types (detached, semi-detached, overcontact, ellipsoidal). The method, related to a more widely used Principal Component Analysis, produces a lower-dimensional representation of the input data while preserving local geometry and, consequently, the similarity between neighboring data points. We use this property to reduce the dimensionality in a series of steps to a one-dimensional manifold and classify light curves with a single parameter that is a measure of "detachedness" of the system. This fully automated classification correlates well with the manual determination of morphology from the data release, and also efficiently highlights any misclassified objects. Once a lower-dimensional 
 projection space is defined, the classification of additional light curves runs in a negligible time and the method can therefore be used as a fully automated classifier in pipeline structures. The classifier forms a tier of the \textit{Kepler} EB pipeline that pre-processes light curves for the artificial intelligence based parameter estimator.
\end{abstract}

\keywords{binaries: eclipsing --- methods: data analysis --- methods: numerical}

\section{Introduction}

The NASA \textit{Kepler} mission carries out essentially uninterrupted, ultrahigh precision photometric observations of $\sim 160,000$ stars in the Cygnus and Lyra constellations since its launch in 2009 March. It is designed to discover exoplanets by the transit method. The details regarding mission operation and instrumentation can be found in \citet{2010ApJ...713L.109B}, \citet{2010Sci...327..977B}, \citet{2010ApJ...713L..92C}, \citet{2010ApJ...713L..79K}. The mission already yielded several exciting results, including the first ever discovery of transiting circumbinary planets \citep{2011Sci...333.1602D, 2012Nature} and the detection of several Earth sized planets \citep{1112.4514,2011arXiv1112.4550F, 2012arXiv1201.2189M}. In addition to its impact on exoplanetary science, \textit{Kepler} literally revolutionized the eclipsing binary (EB) field. A set of 2165 EBs has been identified and made available via two data releases \citep[][parts I and II, respectively]{2011AJ....141...83P,2011AJ....142..160S} and quite a few papers studied individual \textit{Kepler} EBs \citep[e.g.,][]{2011MNRAS.414.2413S, 2011MNRAS.417L..31S,2011ApJS..197....4W,2011ApJ...741L...1W}.

The EB analysis pipeline employed for the construction of the catalog consisted of the following steps: (1) \emph{EB signature detection}: all threshold crossing events (abrupt changes in the source flux) identified by the main \textit{Kepler} pipeline have been tagged and all recurring events considered; (2) \emph{data detrending}: all intrinsic variability (such as chromospheric activity, interaction, etc.) and extrinsic variability (i.e., third light contamination and instrumental artifacts) are removed by the iterative fitting of the photometric baseline; (3) \emph{the determination of the ephemerides}: the time-space data are phase folded and the dispersion minimized; (4) \emph{morphological classification}: the determination of the EB morphology based on the light curve shape is done by manual inspection into detached (D), semi-detached (SD), overcontact (OC), ellipsoidal (ELV), and uncertain (UNC) sources; (5) \emph{analytic approximation}: every light curve is fit by a polynomial
  chain ({\texttt polyfit}; \citealt{2008ApJ...687..542P}) to find an analytic description; and (6) \emph{EB characterization}: submitting the analytic approximation to the artificial intelligence based tier EBAI \citep{2008ApJ...687..542P}. The tier consists of a set of \emph{morphology-based} neural networks to perform a nonlinear regression of light curves and obtain the principal parameters of EBs. 

The inadequate part of the pipeline is morphological classification because it is based on visual inspection. This approach is inherently subjective and the classification may be unreliable and is prone to errors. Moreover, this is not a feasible approach for large data sets, so an automated approach is much more favorable. There have already been significant efforts focusing on the topic of automatic classification of variable star light curves \citep[][among others]{2005MNRAS.358...30E,2006A&A...446..395S,2007A&A...475.1159D,2011MNRAS.418...96B,2011MNRAS.414.2602D,2011ApJ...733...10R}. 
Most of the employed methods share a common point of using training sets of known light curves that define the classes to which the unknown light curves are assigned to. They also work for a broader range or variables and include many different pulsating and other types besides eclipsing binaries. Since our goal is to produce a classification of \textit{Kepler} EB light curves based on the similarity between the light curves themselves and not to classify them according to the externally defined classes, we instead propose a method of Locally Linear Embedding (LLE). The method was devised by \citet{2000Sci...290.2323R} as a general nonlinear dimensionality reduction tool. It already found its application in astronomy and has recently been used for classification of Sloan Digital Sky Survey galaxy and quasar spectra \citep{2009AJ....138.1365V} as well as stellar spectra \citep{2011AJ....142..203D}.

%To overcome this drawback, we propose a classification of \textit{Kepler} EB light curves by an automated method of \emph{Locally Linear Embedding} (LLE). 

\section{Eclipsing Binary Data}

The data classified in this paper are a part of the second \textit{Kepler} catalog of EB stars \citep{2011AJ....142..160S}. The catalog contains 2165 EBs from the first three quarters (Q0, Q1, and Q2). All light curves were processed and calibrated with the \textit{Kepler} EB pipeline. According to the manual classification, the sample consists of 1261 detached, 152 semi-detached, 469 overcontact binaries, 137 ellipsoidal variables, and 146 uncertain systems.

For our analysis, the phase-folded light curves with their primary eclipse centered at the zero phase were fit using \texttt{polyfit}. In general, \texttt{polyfit} uses a chain of $m$ piece-wise connected polynomials of order $n$ to approximate light curves ranging from well detached systems with narrow primary eclipses to overcontact systems with near-sinusoidal signals. In particular, we fitted a chain of four quadratic functions. Note that this choice is not motivated by any EB physics but is rather the simplest model that performs reasonably well.

Having calculated \texttt{polyfit} models for all EBs, we sampled these analytic representations in 1000 equidistant phase points between $-0.5$ and 0.5. The number of sampling points was selected so that even the narrowest primary eclipses of detached systems are suitably covered.

\section{Locally Linear Embedding}

The method was introduced by \citet{2000Sci...290.2323R} as a dimensionality reduction alternative to the more commonly used algorithms like the Principal Component Analysis (PCA) or Multidimensional Scaling. A particular appeal of this method is in the determination of the \emph{local} relations between data points rather than global properties of the data set.

Considering the remarkable effectiveness of the method, its implementation is relatively simple and can be outlined in three main steps. A detailed description and derivation can be found in \citet{2000Sci...290.2323R}, \citet{2002dRD}, \citet{2003SR} and \citet{2009AJ....138.1365V}. We recap the main steps here.

Consider each \texttt{polyfit} light curve sampled at 1000 equidistant phases as a point in a $D=1000$ dimensional phase space, where the number of dimensions is determined by the number of phase bins in which we sample the model light curve. We aim to represent the data from this high-dimensional space in a significantly lower-dimensional space (say, $d=2$ or 3) where correlations between light curves can be more easily understood and visualized. In the \texttt{polyfit} space, all light curve models are represented by $N$ vectors $\mathbf{x}_i$. In the first step we identify $k$ nearest neighbors per data point, $\mathbf{x}_j$, $j=1\dots k$ for every $i$, as defined by the Euclidean distance. This particular choice of the metric does not prefer any of the dimensions since all of them contribute equally to the distance, however the metric can generally be non-Euclidean.

Next, the local geometry of each data point is characterized by a linear combination of its neighbors. This is appropriate only if the neighborhood points are always close enough to each data point so that the linear approximation is valid. The cost function that measures the reconstruction error is written as
\begin{equation}\label{eq:cost_func}
\mathcal{E}(W)=\sum_i   \Bigl| \textbf{x}_i - \sum_j w_{ij}\textbf{x}_j \Bigr|^2,
\end{equation}
where the inner summation goes over all of the neighbors of each data point $i$, calculated in the previous step, and the outer summation goes over all data points. Weights $w_{ij}$ describe the contribution of the $j$th neighbor to the reconstruction of the $i$th point. We seek such weights that, when multiplied by the vectors of the neighborhood points, will minimize the difference between the $i$th point and this weighted sum. The weights of interest may be computed by minimizing the cost function, subject to the constraint:
\begin{equation}\label{eq:weights}
\sum_j w_j=1.
\end{equation}
This is done by using a standard Lagrangian multiplier method. By using Equation~(\ref{eq:weights}), the $i$th component of Equation~(\ref{eq:cost_func}) can be rewritten as
\begin{equation}
\mathcal{E}^i(W)=\Bigl| \sum_j w_{ij}(\textbf{x}_i-\textbf{x}_j) \Bigr|^2=\sum_{j=1}^k\sum_{l=1}^k w_{j}w_{l}C_{jl},
\end{equation}
where
\begin{equation}
C_{jl}=\left(\textbf{x}_i-\textbf{x}_j\right)^T \left(\textbf{x}_i-\textbf{x}_l\right)
\end{equation}
is the neighborhood correlation matrix. The optimal weights are then given by
\begin{equation}
w_j=\frac{\sum_l C_{jl}^ {-1}}{\sum_{m}\sum_{n}C_{mn}^{-1}}.
\end{equation}
Instead of computing an inverse of the matrix $\mathbf{C}$, a more practical approach is to solve the linear system:
\begin{equation}
\sum_j C_{jl}w_l=1,
\end{equation}
and rescale the weights so that they satisfy the condition in Equation~(\ref{eq:weights}). In general, the matrix $\mathbf{C}$ can be singular and so the weights $w_j$ would be ill defined. A way to overcome this problem is to add a small multiple of the identity matrix to the correlation matrix:
\begin{equation}
\mathbf{C}\to \mathbf{C}+r\mathbf{I}.
\end{equation}
\citet{2009AJ....138.1365V} discovered that the value of the regularization parameter $r=10^ {-3}\mathrm{Tr}(\mathbf{C})$ works well.

A final, third step of the method involves mapping of the data points onto a lower, $d$-dimensional space $Y$. The embedding cost function that needs to be minimized in order to find the projected space $Y$ can be written in a similar fashion as Equation~(\ref{eq:cost_func}):
\begin{equation}\label{eq:cost_func_y}
\mathcal{E}(Y)=\sum_i \Bigl| \textbf{y}_i - \sum_j w_{ij}\textbf{y}_j \Bigr|^2.
\end{equation}
Instead of optimizing the weights $w_{ij}$, we keep them fixed and optimize the coordinates $\mathbf{y}_i$. This way we find a low-dimensional space whose local patches are geometrically equivalent to the patches of the high-dimensional space, meaning that the intrinsic relations between neighboring data points are (approximately) preserved. The computation of the optimal space $Y$ reduces to a calculation of the eigenvectors of a sparse matrix (for details, see one of the listed sources for LLE). The projection is defined by the first $d$ eigenvectors that correspond to the smallest eigenvalues and the solution with the zero eigenvalue (which exist because of the constraint in Equation~(\ref{eq:weights})) and can be omitted since it only amounts to the translation in space \citep{2000Sci...290.2323R}.

Once the low-dimensional space $Y$ is defined, new data points are easily projected onto it by finding their nearest neighbors and reconstructing the weights. After that only the projected coordinates need to be calculated from the absolute value term in Equation~(\ref{eq:cost_func_y}).

The advantage of LLE over other dimensionality reduction methods is in its low number of free parameters. When the input data are defined, the only two free parameters are the number of nearest neighbors, from which the local reconstruction is made, and the regularization parameter $r$. The algorithm is  not computationally intensive and the classification procedure of the whole sample takes only a few seconds on a personal computer.

\section{Classification Results}

\begin{figure}
\begin{center}
\includegraphics[width=0.8\textwidth]{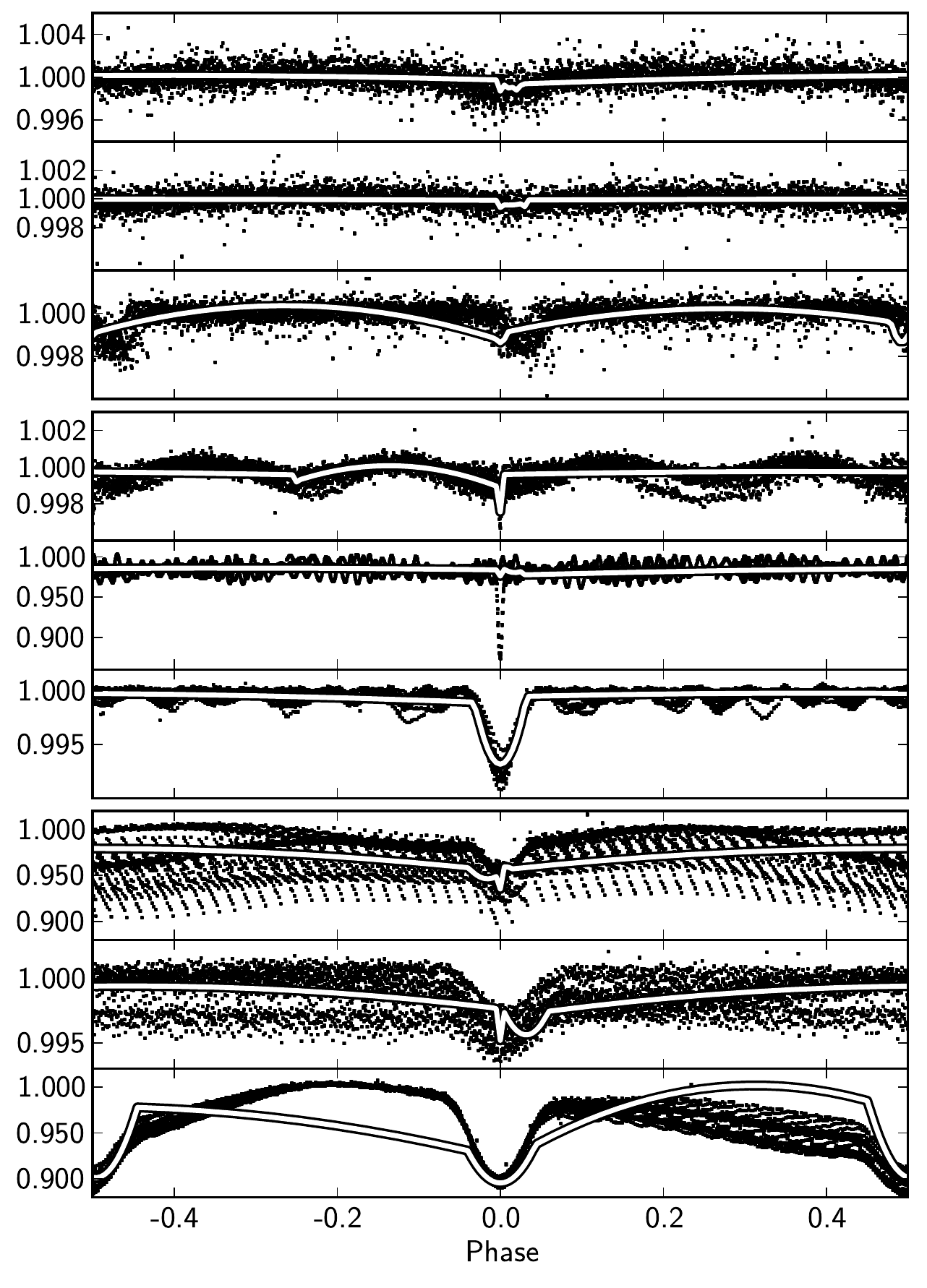}
\end{center}
\figcaption{\label{fig:outliers} Three different groups of outliers as described in the text. The black dots are observed data, and the white lines are \textit{polyfit} light curves. The top three examples show light curves with small eclipse depth; the next three light curves feature additional intrinsic variations; and the last three light curves suffer from bad detrending or bad fits. In all cases \textit{polyfit} failed to adequately represent the underlying observed light curve.}
\end{figure}

The \textit{Kepler} EB catalog includes 2165 objects. Of those, 42 have longer periods than the span of \textit{Kepler} observations and, therefore, the \texttt{polyfit} model could not be reliably fitted. Those objects were omitted from further analysis, leaving 2123 EBs.

All input light curves were then vertically rescaled to the unit ($[0,1]$) interval. This way, the effects of the third light contamination and orbital inclination are minimized. Had this not been done, the depth of the eclipses would have been the driving parameter and classification would have been dominantly triggered on the overall amplitude rather than on the morphology of the light curve. Given the large {\it Kepler} pixel size ($4'' \times 4''$), a significant number of targets is contaminated by third light and this causes additional problems for classifying un-normalized data. If we used LLE on the non-scaled data, the light curves would be classified by amplitude; hence, any detached, semi-detached, or overcontact systems with similar eclipse depths would be lumped together, while two morphologically identical sources, one diluted by third light and the other undiluted, would end up in the different parts of the LLE manifold.

\begin{figure*}
\plotone{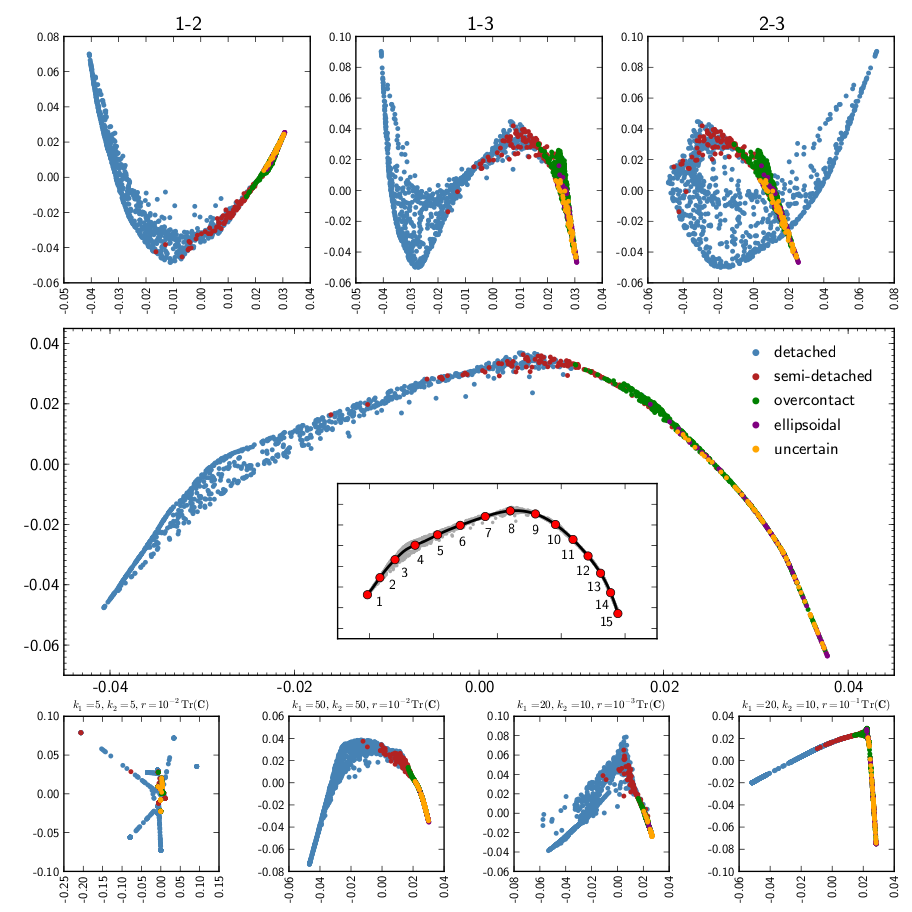}
\figcaption{\label{fig:lle} LLE projection of 1572 \textit{Kepler} EB light curves sampled in 1000 phase points $(D=1000)$ and first projected onto $d_1=3$D subspace (the upper three panels show individual component cross-sections), followed by a projection to a $d_2=2$D subspace. Note that the values on the $x$ and $y$ axes of the diagrams are arbitrary since the projections do not depend on global translations, rotations and scalings. EB classes from existing manual classification are depicted by different colors. The inset in the middle diagram shows the spline fit to the main line and a selection of 15 light curves depicted in Figure~\ref{fig:lcs}. The bottom four diagrams show the LLE projection using different values for the number of nearest neighbors used in the calculations and the values of the regularization parameter $r$. For all four cases $k_1$ is the number of nearest neighbors used in the first projection, and $k_2$ is the number used in the second projection.}
\end{figure*}

Initially we calculated the projection for the whole data set; we varied the number of $k$ nearest neighbors for each data point and the regularization parameter $r$. Regardless of the choice of those two parameters (within reasonable limits), the projection always highlighted a group of outliers. Closer inspection of those objects revealed that \texttt{polyfit} failed to adequately fit their observed light curves, mainly for three reasons: (1) small eclipse depths ($\sim 10^{-3}$) are buried in noise; (2) the presence of intrinsic variations (spots, pulsations, accretion, \dots); and (3) difficulties with detrending (Figure~\ref{fig:outliers}). In some cases the affected light curves could be represented with a larger number of higher order polynomials, but this would more likely lead to overfitting rather than giving a better representation. For that reason we excluded 551 affected light curves from the sample used to generate the mapping. After pruning, 1572 EBs were used to
  compute the final projection. This does not affect the final classification sample, as all sources, including any subsequently added ones, can still be mapped to the lower-dimensional space by the same projection.

\begin{figure*}
\plotone{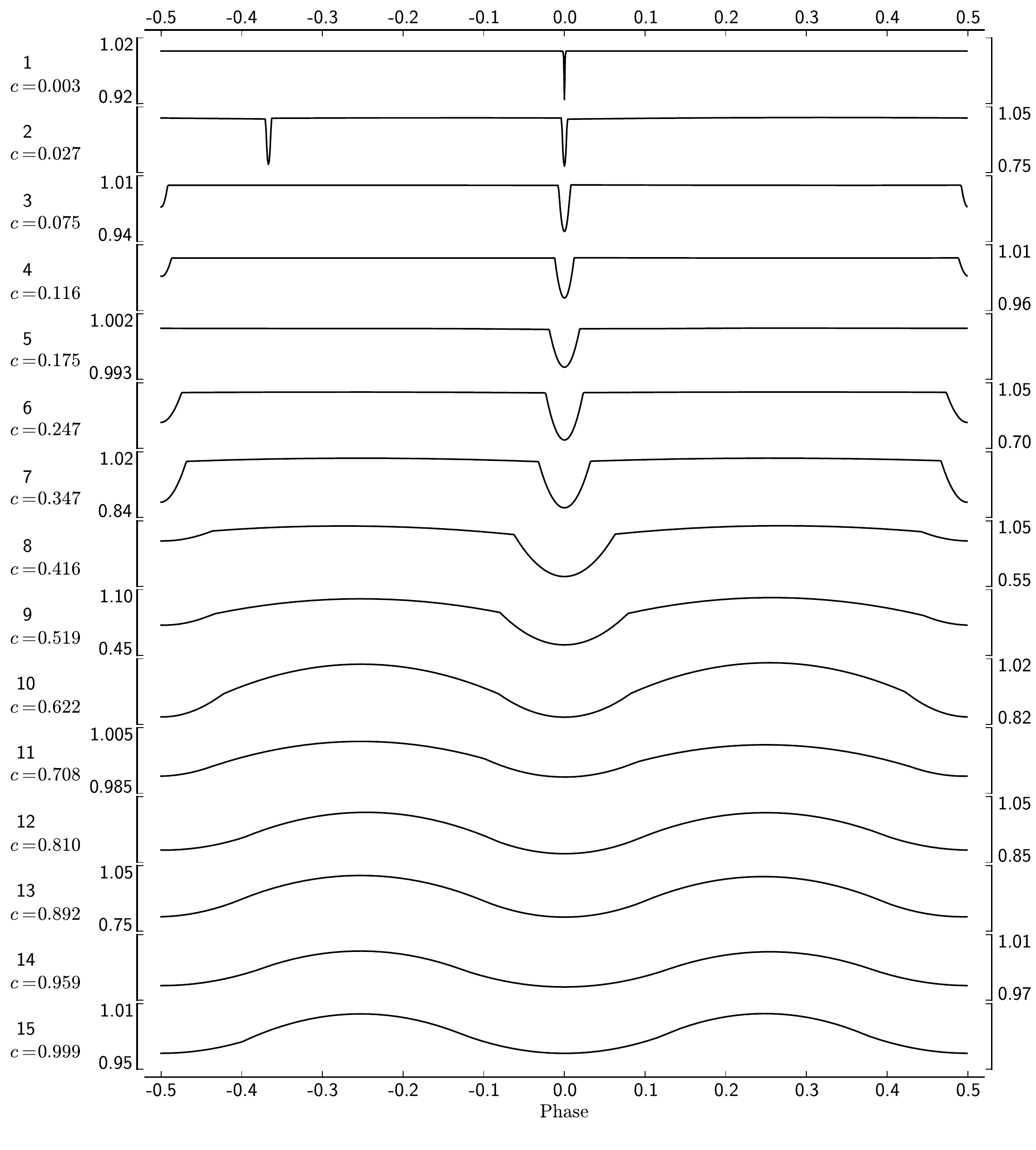}
\figcaption{\label{fig:lcs} Polyfit representations of EB light curves along the main classification sequence on the two-dimensional projection depicted in Figure~\ref{fig:lle}. The light curves are enumerated according to the labeled points on the classification sequence and the classification parameter $c$ is given for each point on the left. The light curves are not rescaled to the $[0,1]$ range as they were for the LLE projection but are rather given in the original scaling. The values along the $y$ axes specify their flux variation ranges.}
\end{figure*}

Experimentation with the final sample showed that the best choice for the number of nearest neighbors is around $k=20$, and that the projection onto a two-dimensional subspace preserves a sufficient amount of morphological information for classification purposes. An example from \citet{2003SR} shows that the exact number of neighbors is not critical; our results confirm that since a modest variation in this number (i.e.,~between 15 and 25) did not affect the projection significantly. For the value of the regularization parameter we chose $r=10^{-2} \, \mathrm{Tr}(\mathbf{C})$, which is somewhat larger than suggested in the literature. The rationale for this is that larger values penalize large weights, producing a smoother projection. The final projection to $d_2=2$ was made in two steps. First, we projected all light curves onto a three-dimensional subspace (cf.~Figure~\ref{fig:lle}, top panels), from where it was evident that the projection is a well-constrained 2-dimensional manifold. We then made another projection from $d_1=3$ to $d_2=2$ with $k=10$ nearest neighbors, depicted in the middle panel of Figure~\ref{fig:lle}. This two-step method yielded a better separation between different classes of EBs than a direct projection to 2 dimensions. Examples of the final projections when calculated with a different set of free parameters are shown in the bottom panel of Figure~\ref{fig:lle}.

The middle panel in Figure~\ref{fig:lle} demonstrates that the projection of light curves can be further represented by the one-dimensional curve with a relatively limited scatter. In a good portion of all cases the scatter is caused by the differences in the flux levels and phase positions of secondary eclipses. The density of points along the main classification sequence is roughly constant. There is a remarkable correspondence between the LLE determination of morphology and the manual EB classification flag (colors levels in Figure~\ref{fig:lle}). There is, however, some overlap between the manually classified types, most notably between detached and semidetached systems, and between overcontact, ellipsoidal, and uncertain systems. Closer inspection reveals that this overlap is real and that the points that overlap are indeed intrinsically similar. This clearly implies that manual classification suffered from subjective notions of the classifier. Morphology types are thus aligned
with the main classification sequence: the leftmost part of the sequence is populated by well detached systems with very narrow primary eclipses, followed by the detached systems with progressively wider eclipses that finally morph into semidetached systems. These further advance into a region of overcontact systems that ends with an overlap of population of systems exhibiting near-sinusoidal variations. Curiously, the rightmost part of the sequence is where all the unknown/uncertain systems reside, implying that the dubious classification stems from light curve similarity that cannot be resolved without additional data, i.e.,~follow-up spectroscopy.

In order to quantitatively characterize different classes of EBs, we fit the main classification sequence with a smoothed spline function (cf.~the middle panel in Figure~\ref{fig:lle}). We define a classification parameter $c$ for each light curve along the fitted spline that is closest to a given data point, ranging from 0 to 1. A selection of 15 light curves that lie along the spline is shown in Figure~\ref{fig:lcs}. The width of the primary eclipse changes continuously with the classification parameter $c$; the depth and the phase of the secondary eclipse do not influence the classification notably. Since the width of the primary eclipse is the measure of the sum of relative radii of the two components, the classification parameter is a good measure of the relative separation between the components in the binary system. However, it should be noted that EB physics is too complex to be put in only a single parameter so the values of the classification parameter should only be used as guidelines.

\begin{figure}
\plotone{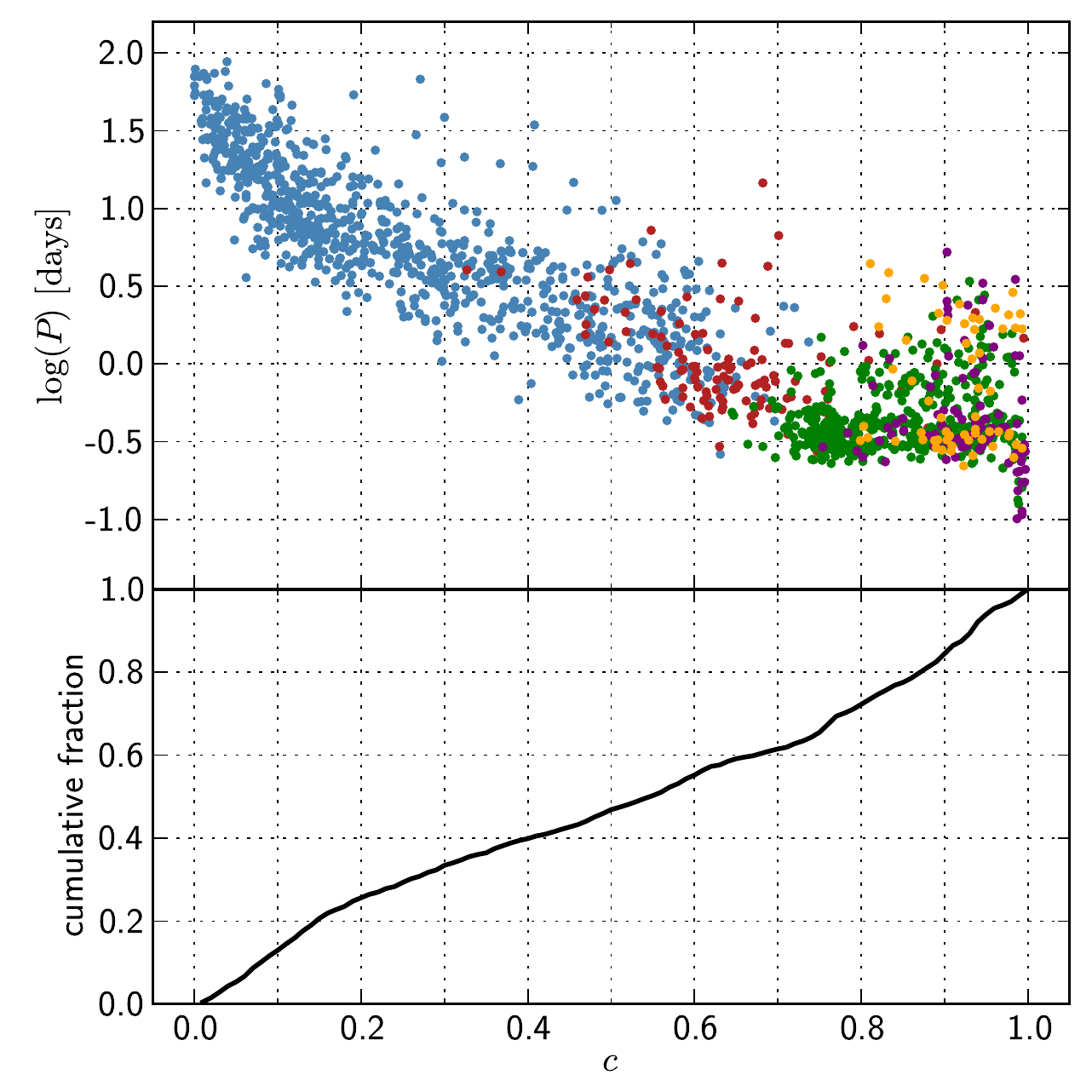}
\figcaption{\label{fig:period} Correlation between the orbital period $P$ and the classification parameter $c$. Colors correspond to the manually classified EB types as in Figure~\ref{fig:lle}. The bottom diagram shows the cumulative histogram with respect to $c$.}
\end{figure}

Since the classification parameter and the sum of relative radii are notably correlated, we tested if correlation exists between the orbital periods and $c$. The obvious trend is depicted in Figure~\ref{fig:period}: the longer period systems, which are predominantly detached, have small values of the classification parameter. Overcontact systems and, more so, ellipsoidal variables are clumped in the region with short periods and large values of $c$. The values of $c$ that separate different EB classes could be estimated by comparing them to the manual classification. As a guideline, all systems with $c<0.5$ are predominantly detached. The range of $c$ for semidetached systems is harder to estimate due to the overlaps, but broadly they lie in the $0.5<c<0.7$ range. Overcontact systems are dominating $0.7<c<0.8$ region, after which a mixture of ellipsoidal variables and systems with uncertain classification sets in.

\section{Update to the catalog}

The live version of the \textit{Kepler} EB catalog is served on \verb|http://keplerEBs.villanova.edu| and on MAST, \verb|http://archive.stsci.edu/kepler|. With this paper we introduce a new column to the catalog that contains the classification parameter $c$. The community can thus further refine the mining of the catalog according to the particular scientific interest. For example, a search can be limited to well-detached EBs by querying the entries with $c < 0.1$ or to near-sinusoidal light curves for the entries with $c > 0.9$.

\section{Summary}

In this paper we present a \emph{fully automated}, objective \textit{Kepler} EB light curve classification scheme based on the LLE algorithm. The method highlights the local geometrical properties of light curves in high-dimensional space (spun by the number of phase points in a light curve) and projects them to a lower-dimensional subspace. For our purposes we projected the light curves onto a two-dimensional space where individual morphological types are well defined by a spline fit with relatively little scatter. Based on that, we define a classification parameter $c$, a single parameter that marches along the spline and, hence, provides a quantitative representation of the binary star morphology.

The classification of the whole sample runs in negligible time; once the projected subspace is defined, additional light curves can be classified on the fly (several thousand light curves per second on a typical personal computer). This fact makes our approach appealing for integration into the existing analysis pipelines.

However, since the classification of a light curve depends on the automated \texttt{polyfit} model, the latter should always be inspected for any systematic problems. The classification parameter readily serves as a "best guess" at the morphology class, but full-fledged modeling might still be necessary to confirm the classification reliability.

\acknowledgements
Funding for the \textit{Kepler} mission is provided by NASA's Science Mission Directorate. G.M. acknowledges support from the Slovenian Research Agency. A.P. acknowledges the {\it Kepler} Participating Scientist Program NSR 303065. S.B. acknowledges funding from the European Research Council under the European Community's Seventh Framework Programme (FP7/2007--2013)/ERC grant agreement no. 227224 (PROSPERITY), as well as from the Research Council of K. U. Leuven grant agreement GOA/2008/04.

\end{document}